# Elicitation and Modeling Non-Functional Requirements – A POS Case Study

Md. Mijanur Rahman and Shamim Ripon, *Member IACSIT*

*Abstract*— Proper management of requirements is crucial to successful development software within limited time and cost. Nonfunctional requirements (NFR) are one of the key criteria to derive a comparison among various software systems. In most of software development NFR have be specified as an additional requirement of software. NFRs such as performance, reliability, maintainability, security, accuracy etc. have to be considered at the early stage of software development as functional requirement (FR). However, identifying NFR is not an easy task. Although there are well developed techniques for eliciting functional requirement, there is a lack of elicitation mechanism for NFR and there is no proper consensus regarding NFR elicitation techniques. Eliciting NFRs are considered to be one of the challenging jobs in requirement analysis. This paper proposes a UML use case based questionary approach to identifying and classifying NFR of a system. The proposed approach is illustrated by using a Point of Sale (POS) case study.

## I. INTRODUCTION

The process of discovering, documenting, analyzing, and checking the constraints and service is called the requirement engineering [1]. Requirement Engineering is one of the major areas of software engineering. It is an early step for system development. System development can be successful only with consistent requirement management. Quality requirement has a huge impact on the final product [2], [3]. Requirements are categorized into Functional and Nonfunctional requirements. Functional requirements describe the external and internal visible output of a system [4]. Nonfunctional requirements, on the other hand, are the constraints of the system. These constraints are for development and deployment process. The quality requirements are also known as nonfunctional requirement [5]. The particular quality the system must have like accuracy, performance, usability, modifiability, safety, performance, reliability, security, flexibility, etc. [6]. NFR are always connected with a functional requirement [7].

Unfortunately still now system analysts are not very much aware of nonfunctional requirements. Where functional requirements are gathered at an early stage of system development, ignorance of nonfunctional requirement can lead to project failure. A common problem is that very often stakeholders are not aware of their NFR requirements. It is very hard for the stakeholders to know the details about NFRs.

The rate of project failure is increasing because of insufficient NFR gathering at the proper stage. NFRs have been treated as the properties or attributes of system which is needed to satisfy the customers. In many cases customers' expectation are not fulfilled because of inadequacy of the system properties. The cost and time to market of software development can be reduced by giving more importance on nonfunctional requirement. Customers do not know the constraints of system in the early stage of the development process. Even the system developer does not focus on the NFRs at the beginning of system development. In a complex system, NFRs are vital and sensitive. The system can be threatened if NFRs are neglected during the system development. Since the complexity of software is increasing and customers are focusing more on quality of software, NFR is no longer considered a secondary option in requirement elicitation process. For these reason, it is required to focus on eliciting and modeling of NFRs.

Although there are standard definitions of functional requirements, there is a lack of well-formed definition of NFR. To formally specify and characterize the NFRs are very much harder [6], because NRFs vary in different circumstances. Sometimes both functional and nonfunctional requirements are mixed up and ambiguity arises differentiating between them. Since nonfunctional requirements are linked with functional requirements, they create conflicts among stakeholders, e.g., security of a system can be two level password or biometric system, but the later will increase the cost of the product which is associated with nonfunctional requirement. For the lack of domain knowledge we do not get adequate NFRs, besides it is not even certain which NFR will be taken into consideration. NFR is not equally considered as functional requirements in software development.

Requirement gathering or discovering is known as the elicitation process. Elicitation is one of the crucial issues for the system development and a major part of the requirement engineering. In software development process, one of the most critical knowledge-intensive activities is requirement elicitation [8]. NFRs are prioritized from stakeholder's point of view [9], so it should be first elicited from the stakeholders. So the elicitation technique has to be designed in such a way that it will interact closely with the stakeholders. Formal technique such as UML use case models is very useful for discovering FRs [10]. One of the major activities of requirement engineering is requirement elicitation and Md. Mijanur Rahman is with East West University and Daffodil International University, Dhaka, Bangladesh (e-mail: mmr.sinha@yahoo.com).

Shamim Ripon is with the Department of Computer Science and Engineering, East West University, Dhaka, Bangladesh (e-mail: dshr@ewubd.edu).

analyzing [1]. There is no proper elicitation method available for NFRs. Apart from giving formal notations [4], recently a few elicitation techniques have been proposed with some strength and weakness. But still the requirement engineering community is not yet agreed to define any of the technique as a standard. A major problem of NFR is how to measure the NFR and how to deploy nonfunctional requirement. Representation and elicitation are crucial challenges for NFR. Most of the time conflicts arise among stakeholders because of NFR.

Only a few elicitation techniques have been proposed to discover NFRs. Few of these approaches suggested integrating NFR with FR to avoid bigger maintenance cost [6], [1]. In this paper we focus on NFR discovering and categorizing. We extend UML use case diagram facilitating both eliciting and modeling NFRs where our proposal is based on questionnaires approach to identifying NFR. We illustrate our approach by eliciting and modeling NFRs of a Point of Sale (POS) system.

In the rest of the paper, Section II gives a brief overview of POS system illustrating various NFRs of the system. Section III describes the NFR elicitation process by extending the use case diagram and tabular based questionnaires. Finally, we conclude our paper and outline our future plans in Section IV.

## II. POINT OF SALE (POS) OVERVIEW

Point of sale is a terminal or physical location where goods are sold to customers and exchange transactions. These sales and transactions are occurred by using a computer system called point of sales (POS) system. A POS system is a software which runs on computer. Buy and sales occurs using software (POS) without paper calculation. A POS system does not use traditional system for transaction; it is an automated system. A POS system consists of a computer, barcode scanner, cash drawer, receipt printer and the POS software. In a traditional system for managing sales there is a need for some employees. Training is needed for the employees to process sales and it cost additional investments. Most of the time, it is not possible to manage skilled sales persons for this purpose. Hence, POS system becomes an evident solution. A POS system handles the final transaction and sales, and calculates the total by tracking every sell. POS software integrates inventory, account receivable in real time. It checks credit limit of customer's account. It also supports authorization and processing of debit and credit cards. A POS system is a combination of software, hardware and peripheral devices. A traditional POS system is depicted in Fig.1.

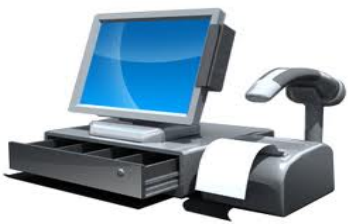

Fig.1 Point of Sale (POS) System

Point of sale is one of the important parts of business now days. POS reduces the paper works, provides a better control of operation and increases the efficiency of daily operation. POS system can save money and increase productivity. It increases accuracy and profit margin, cut expenses, improves service. In sales, times are spent by gathering sales figures or other repetitive work. Important paper work kill most of times like tax reporting, payroll, inventory control, sales monitoring, sales reporting, payment reporting. With a few keystrokes a POS is able to get detailed information daily, weekly, monthly and yearly. Usually the stakeholder of POS is customer, Sales man, Cashier, Manager, Administrator and staff. The processing steps in a POS system are shown in Fig.2.

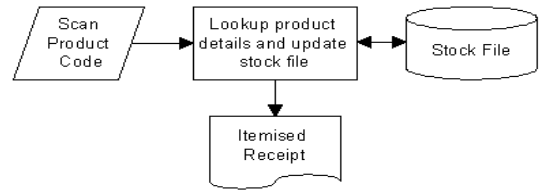

Fig.2 Processing steps of a POS

After analyzing a POS system the following features are identified:
1) It has attached with other devices like barcode, scanner, keyboard, printer, payment terminal (for debit/ credit cards), displays for the customer.
2) POS system maintains an audit trail which contains tracks of the transactions performed within the system.
3) Super admin can create, update, delete and read people role and action
4) Authorized user can create, Delete, update, and read products
5) Authorized user can create Delete, update, read order and payment
6) Authorized user can Add, Delete, and update product information
7) User can Add, and update their information
8) Authorized user can view report.
9) POS cashier can access to customers information
10) POS cashier can search product
11) POS cashier handle payment
12) Salesman process sells
13) System generate barcode

## III. NFR ELICITATION

### A. NFR Process Model

In our proposed model UML use case is used to depict FR. After collecting the FR we draw the use case diagrams. In the use case diagram, questionnaires are integrated with functional requirement. Since NFRs are linked with FR, we add possible questions with each FR of use case. We then get the answers of the questions which is our expected NFRs. For one FR can be multiple questions which are numbered as NFRQ1, NFRQ2,..,NFRQn, where question numbers are written in dotted diamond box and questions are written in the dotted rectangular box. Association between functional requirement and related NFR questions is drawn by using dotted lines. Fig. 3 illustrates a part of POS use case describing NFR elicitation question at various stages of FR.

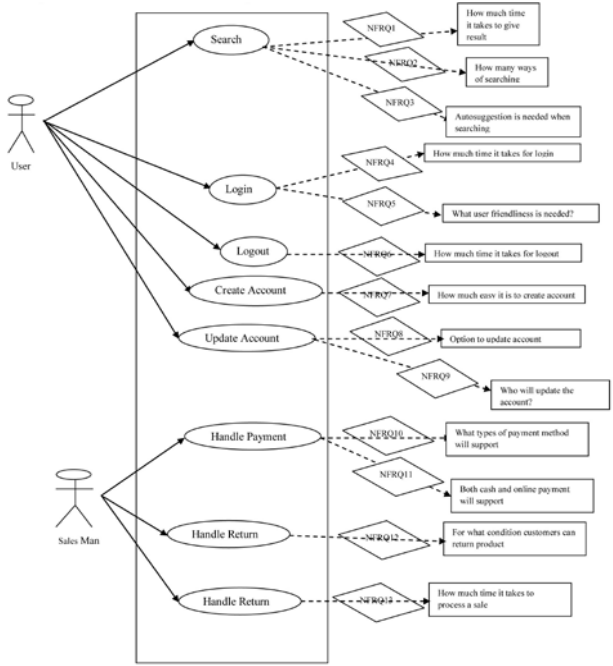

Fig. 3 NFR elicitation questions in use case diagram

We not only elicit the NFRs but also identify the categories of NFRs. Each elicited NFR are categorized into a set of previously well-defined NFR categories. Such categorization will later facilitate modeling and tracking NFR at various stages of system development. Fig. 3 and Fig. 4 illustrate how elicited NFRs are categorized in use case diagrams.

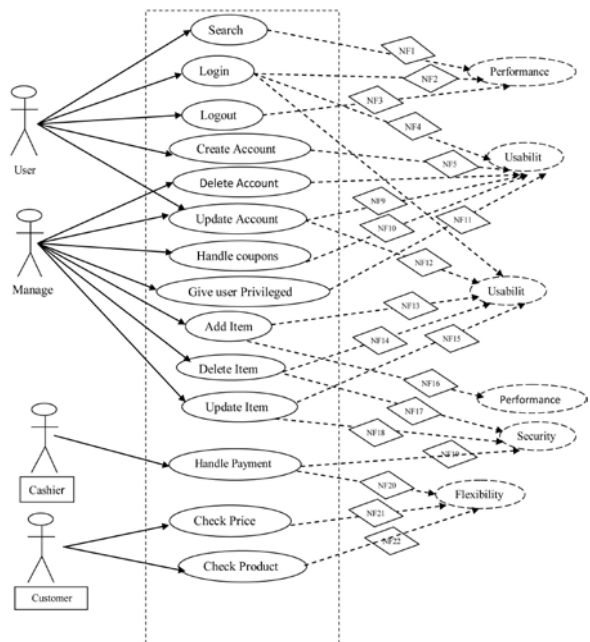

Fig. 4 Categorization of elicited NFR (partial)

In addition to use cases, we also use tabular representation, where the actors, functional requirements, question no, NFR questions and answer and categories of NFR are presented column wise.

### B. NFR Categorization:

Usually, developers collect functional requirements from the very early stage of system development and draw the use diagrams of the system. From the above elicitation technique we just add the possible question to the functional requirements where the answers come from the stakeholders and the given answers are the NFRs. We ask possible question which is linked with functional requirements in use case. In the table, we write the answer of the questions to get NFRs and classify them into predefined categories. For example, in Fig.3, one of the actors is User who has a search function and for this FR we can ask NFRQ2: "how many ways are available for searching?" We can get an answer and it is under Flexibility category. Table 1 illustrates part of the elicited and categorized NFRs.

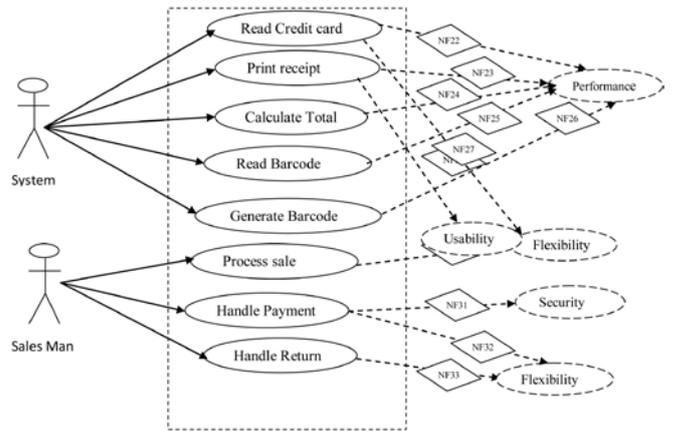

Fig. 5 NFR categorization (partial)

### C. NFR Checklist

After eliciting and categorizing the NFR, a checklist is developed based one the available FR and the elicited NFR against each FR. We only consider the widely used NFR categories against each NFR in this paper. The checklist is shown in Table 2.

## IV. CONCLUSION

Acceptance of any software depends on the customer satisfaction which largely depends on maximizing NFR elicitation and incorporation in the software product. In this paper we have proposed NFRs elicitation technique based on use case extensions. We have extended use case diagram and incorporate NFRs eliciting questions with functional requirements. We have also illustrated an extension of UML use case diagram to model the NFRs that also facilitates the elicitation of NFRs. This technique is based on asking queries for nonfunctional requirements which are available in use case and answer will be collected from stakeholder. We have also categorized the elicited NFRs. In this work in-progress paper we have elicited NFRs with a systematic approach of a system which is Point Of Sale system. This requirement elicitation technique spans from NFRs elicitation to its categorization and finally, showed a check list of widely used NFR that are addressed in our proposed mechanism. Using our NFR elicitation technique we can identify most of the commonly used NFRs. The tabular representation is helpful for keeping track of the NFRs at the various levels of requirement in a system. It is convenient to understand both developer and customer which are less cost effective.

TABLE 1 CATEGORIZING AND ARRANGING NFRS (PARTIAL)

| Actor/ Stake holder | Use Case (Functional Requirement) | Question no | Question for NFR | Question Answer (Elicited NFR) | Category of NFR |
|---|---|---|---|---|---|
| User | Search | NFRQ1 | How much time it takes to give Search result | Less than 10 second | Performance |
| User | Search | NFRQ2 | How many ways of searching | Full and partial match word | Flexibility |
| User | Search | NFRQ3 | Autosuggestion is needed when searching | When writing for searching show related work | Usability |
| User | Login | NFRQ4 | How much time it takes for login | Less than 30 sec | Performance |
| User | Login | NFRQ5 | What is the user friendliness needed | Show message if submit without user name or password | Usability |
| User | Logout | NFRQ6 | How much time it takes for logout | Less than 30 second | Performance |
| Use | Create Account | NFRQ7 | How much easy it is to create account | Use drop down box to select relevant option | Usability |

TABLE 2 CHECKLIST FOR NFR ELICITED NFR

| NFR → FR ↓ | Performance | Flexibility | Usability | Modifiability | Privacy | Legal issue | Security |
|---|---|---|---|---|---|---|---|
| Search | ✓ | ✓ | ✓ | | | | |
| Login | ✓ | | ✓ | | | | |
| Logout | ✓ | | | | | | |
| Create Account | | | ✓ | | | | |
| Update Account | | | | ✓ | ✓ | | |
| Handle Payment | | ✓ | | | | ✓ | |
| Process Sale | ✓ | | | | | | |
| Delete Account | | | ✓ | | | | ✓ |
| Handle Coupon | | | | | | ✓ | |
| Add Item | ✓ | ✓ | | | | | |
| Delete Item | | ✓ | | | | | |
| Update Item | | | ✓ | | | | |
| Give User Privileged | | | | | | | ✓ |
| Read Credit Card | ✓ | | ✓ | | ✓ | | |
| Print Receipt | ✓ | | ✓ | | | | |
| Read Barcode | ✓ | | ✓ | | | | |
| Generate Barcode | | | | | ✓ | | ✓ |
| Calculate Total | ✓ | | | | | | |
| Check Price | | ✓ | | | | | |
| Check Product | | | ✓ | | | | |

A tabular representation has been given with the NFRs and its category. The case studies showed that our technique gave a guidance to elicit sufficient NFRs of a system. The elicited NFRs are measure and traceable because of check list. In the case study there is small chance to elicit irrelevant NFRs but elicited in an easy and structured way. Our future plan includes modeling case studies of other real-life application to experiment the applicability and scalability of the proposed approach. Our proposed approach can be conveniently adapted to software product line to elicit NFRs of system families Based on our earlier experience with software product line [11], [12] we are also planning to model NFR of SPL.

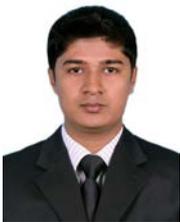
**Md. Mijanur Rahman** is a Lecturer in the Department of Software Engineering, Daffodil International University, Dhaka, Bangladesh. His research interests are in Software Engineering, E-Commerce, E-Learning and Internet Marketing in general. He passed B.Sc. in Computer Science and Engineering from East West University. He is the prominent IT speaker and writer in the Bangladesh as well as ICT consultant in the Syntech Limited. His current research is focused on modeling non functional requirement in the field of requirement engineering.

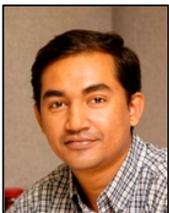
**Shamim Ripon** is an Assistant Professor in the Department of Computer Science and Engineering, East West University, Dhaka, Bangladesh where he leads Software Engineering and Formal Method Research Group. Previously, he was a Research Associate in the Department of Computing Science, University of York, UK and Research Fellow in the Department of Computing Science, University of Glasgow, UK. He also served as a Lecturer in Khulna University, Bangladesh. He is a member of IAENG.

Dr. Ripon holds a B.Sc. in Computer Science and Engineering from Khulna University, MSc in Computer Science from National University of Singapore and PhD in Computer Science from University of Southampton, UK. His research interests focus on the Requirement Engineering, Software Product Line, Semantic Web, Natural Language Processing. His current research examines the formal representation and verification of knowledge based requirement specification.